\documentclass[letterpaper,10pt,conference]{IEEEtran}
\usepackage{graphicx,latexsym,epsfig,amssymb,amsmath,subfigure}

\begin{document}

\title{The Impact of Secure OSs on Internet Security\\ \emph{What Cyber-Insurers Need to Know}}



\author{\authorblockN{Ranjan Pal}
\authorblockA{Department of Computer Science\\
University of Southern California\\
Email: rpal@usc.edu}
\and
\authorblockN{Pan Hui}
\authorblockA{Deutsch Telekom Laboratories\\
Berlin, Germany\\
Email: pan.hui@telekom.de}}


%

\small
\maketitle
{\small
\begin{abstract}
In recent years, researchers have proposed \emph{cyber-insurance} as a suitable risk-management technique for enhancing security in Internet-like distributed systems. However, amongst other factors, information asymmetry between the insurer and the insured, and the inter-dependent and correlated nature of cyber risks have contributed in a big way to the failure of cyber-insurance markets. Security experts have argued in favor of operating system (OS) platform switching (ex., from Windows to Unix-based OSs) or secure OS adoption as being one of the techniques that can potentially mitigate the problems posing a challenge to successful cyber-insurance markets. In this regard we model OS platform switching dynamics using a \emph{social gossip} mechanism and study three important questions related to the nature of the dynamics, for Internet-like distributed systems: (i) which type of networks should cyber-insurers target for insuring?, (ii) what are the bounds on the asymptotic performance level of a network, where the performance parameter is an average function of the long-run individual user willingness to adopt secure OSs?, and (iii) how can cyber-insurers use the topological information of their clients to incentivize/reward them during offering contracts? Our analysis is important to a profit-minded cyber-insurer, who wants to target the right network, design optimal contracts to resolve information asymmetry problems, and at the same time promote the increase of overall network security through increasing secure OS adoption amongst users. 

\end{abstract}
%
\IEEEpeerreviewmaketitle
\section{Introduction}
The Internet has become a fundamental and an integral part of our daily lives. Billions of people nowadays are using the Internet for various types of applications. However, all these applications are running on a network, that was built under assumptions, some of which are no longer valid for today's applications, e,g., that all users on the Internet can be trusted and that there are no malicious elements propagating in the Internet. On the contrary, the infrastructure, the users, and the services offered on the Internet today are all subject to a wide variety of risks. These risks include distributed denial of service attacks, intrusions of various kinds, hacking, phishing, worms, viruses, spams, etc. In order to counter the threats posed by the risks, Internet users\footnote{The term `users' may refer to both, individuals and organizations.} have traditionally resorted to antivirus and anti-spam softwares, firewalls, and other add-ons to reduce the likelihood of being affected by threats. In practice, a large industry (companies like \emph{Symantec, McAfee,} etc.) as well as considerable research efforts are currently centered around developing and deploying tools and techniques to detect threats and anomalies in order to protect the Internet infrastructure and its users from the negative impact of the anomalies.

In the past one and half decade, risk protection techniques from a variety of computer science fields such as cryptography, hardware engineering, and software engineering have continually made improvements. Inspite of such improvements, recent articles by Schneier \cite{sch} and Anderson \cite{ranr}\cite{amr}\cite{ramr} have stated that it is impossible to achieve a 100\% Internet security protection. The authors attribute this impossibility primarily to six reasons: 
\begin{enumerate}
\item Existing technical solutions are not sound, i.e.,there do not always exist fool-proof ways to detect and identify even well  deÞned threats; for example, even state of the art detectors of port scanners and other known anomalies suffer from positive rates of false positives and false negatives \cite{jpbb}. In addition, the originators of  threats, and the threats they produce, evolve automatically in response to detection and mitigation solutions being deployed, which makes it harder to detect and mitigate evolving threat signatures and characteristics \cite{vg}. Other types of damages caused by non-intentional users, such as denial of service as a result of flash crowds, can be predicted and alleviated to some extent but not eliminated completely. Finally, completely eliminating risks would require the use of formal methods to design provably secure systems - however, these methods capture with difficulty the presence of those messy humans, even non malicious humans, in the loop \cite{odlyzko}. 
\item The Internet is a distributed system, where the system users have divergent security interests and incentives, leading to the problem of `misaligned incentives' amongst users. For example, a rational Internet user might well spend \$20 to stop a virus trashing its hard disk, but would hardly have any incentive to invest sufficient amounts in security solutions to prevent its computer being used by an attacker for a service-denial attack on a wealthy corporation like an Amazon or a Microsoft \cite{varian}. Thus, it is evident that the problem of misaligned incentives can be resolved only if liabilities are assigned to parties (users) that can best manage risk. 
\item The risks faced by Internet users are often correlated and interdependent. As a result a user taking protective action in an Internet like distributed system creates positive externalities \cite{hh} for other networked users that in turn may discourage them from making appropriate security investments, leading to the `free-riding' problem \cite{gccr}\cite{jaw}\cite{mybm}\cite{oom}. The free-riding problem leads to suboptimal network security. 
\item Network externalities due to \emph{lock-in} and \emph{first-mover} effects \cite{ranr} affect the adoption of technology. Katz and Shapiro \cite{kschr} have analyzed that externalities lead to the classic S-shaped adoption curve, according to which slow early adoption gives way to rapid deployment once the number of users reaches a critical mass. The initial deployment is subject to user benefits exceeding adoption costs, which occurs only if a minimum number of users adopt a technology; so everyone might wait for others to go first, and the technology never gets deployed. For example, DNSSEC, and S-BGP are secure protocols that have been developed to better DNS and BGP in terms of security performance. However, the challenge is getting them deployed by providing sufficient internal benefits to adopting firms. 
\item Measuring risks is a difficult proposition \cite{pcr}. Risks cannot be managed better or completely eliminated until they can be measured better.
\item Many security software markets have aspects of a \emph{lemons market} \cite{lmr} or even worse, i.e., by looking at security software, even the vendor does not know how secure its software is \cite{ramr}. So buyers have no reason to pay for more protection, and vendors are disinclined to invest time, money, and effort to strengthen their security software code. 
\end{enumerate}

In view of the above mentioned inevitable barriers to 100\% risk mitigation, the need arises for alternative methods of risk management in the Internet. Anderson and Moore \cite{amr} state that microeconomics, game theory, and psychology will play as vital a role in effective risk management in the modern and future Internet, as did the mathematics of cryptography a quarter century ago. In this regard, \emph{cyber-insurance} is a psycho-economic-driven risk-management technique, where risks are transferred to a third party, i.e., an insurance company, in return for a fee, i.e., the \emph{insurance premium}\footnote{Running \emph{vulnerability markets} \cite{ramr} is a reasonable proxy to estimating risks, and in turn helping manage risks better. However, there are certain ethical issues that needs to be resolved in order to properly design such markets. As a result vulnerability markets are not that popular and are non-competitive in nature.}. The concept of cyber-insurance is growing in importance amongst security engineers. The reason for this is three fold: 1) ideally, cyber-insurance increases Internet user safety because the insured increases self-defense as a rational response to the reduction in insurance premium \cite{kmy1}\cite{kmy2}\cite{bs}\cite{yd}. This fact has also been mathematically proven by the authors in \cite{leb3}\cite{leb}, 2) in the IT industry, the mindset of `absolute protection' is slowly changing with the realization that absolute security is impossible and too expensive to even approach, while adequate security is good enough to enable normal functions - the rest of the risk that cannot be mitigated can be transferred to a third party \cite{kmy3}, and 3) cyber-insurance will lead to a market solution that will be aligned with economic incentives of cyber-insurers and users (individuals/organizations) - the cyber-insurers will earn profit from appropriately pricing premiums, whereas users will seek to hedge potential losses. In practice, users generally employ a simultaneous combination of retaining, mitigating, and insuring risks \cite{bs2}.
\subsection{Why Cyber-insurance Has Not Taken Off ?}
Sufficient evidence exists in daily life (e.g., in the form of auto and health insurance) as well as in the academic literature (specifically focused on cyber-insurance\cite{kmy1}\cite{kmy2}\cite{leb3}\cite{leb}\cite{bs} that insurance-based solutions are useful approaches to pursue,i.e., as a complement to other security measures (e.g., anti-virus software).
However, despite all promises, current cyber-insurance markets are non-competitive, specialized, and non-liquid. The inability of cyber-insurance in becoming a common reality is due to a number of unresolved research challenges as well as practical considerations. The most prominent amongst them are \emph{information asymmetry} between the insurer and the insured, and the interdependent and correlated nature of cyber-risks \cite{rmb}\cite{rabohme}. Information asymmetry has a significant effect on most insurance environments, where typical considerations include inability to distinguish between users of different (high and low risk) types, i.e., the \emph{adverse selection} problem, as well as users undertaking actions that affect loss probability after the insurance contract is signed, i.e., the \emph{moral hazard} problem.  However, there are important aspects of information asymmetry that are particular to cyber-insurance for distributed computing environments. These include malicious users hiding information about their intentions and anti-social behavior from their insurers, users lacking information about other networked nodes, as well as insurers lacking information about and not differentiating based on products (e.g., anti-virus software) installed by users. In a recent article \cite{rmb} show that even under ideal conditions of independent and non-correlated cyber-risks, an optimal cyber-insurance contract that accounts for the adverse selection and the moral hazard problem, \emph{overprices} client premiums, in turn de-incentivizing clients to make cyber-insurance markets a success. Thus, it is evident that market-based approaches to mitigate the effects of information asymmetry will most likely fail to enable market success under non-ideal conditions of interdependent and correlated cyber-risks.  
\subsection{A Way Towards Successful Cyber-insurance Markets}
Recently, researchers have argued in favor of \emph{operating system (OS) platform switching} as one technique that could lead to successful cyber-insurance markets in future \cite{rabohme}. A robust OS (ex., Linux based OSs) has more built-in security and permission features than those of its counterparts (ex., Windows) and generally puts its adopters in a \emph{low} risk category \cite{rabohme}, thereby mitigating the adverse selection problem. In addition, a widespread robust OS adoption by web users also reduces the magnitude of interdependent and correlated risks. As far as the moral hazard problem is concerned, effective mechanism design \cite{pg}\cite{plg} will lead to optimal contracts where premiums will seem to be fair to clients, thus promoting markets for cyber-insurance. 

However, not all Internet users adopt the Linux-based OSs like Unix BSD, Linux, and MacOS as their platform of computing. Statistically speaking, most naive Internet users around the world prefer the Windows OS due to its ease of use and its support for myriads of application softwares and web applications \cite{ranr}. Interestingly, the percentage of users adopting MacOS/Unix/Linux as their preferred OS are increasing year after year \emph{(http://www.w3schools.com)}\footnote{A statistical survey reveals that approximately 84\% of Internet users in 2011 are currently adopting Windows OS platforms compared to 90\% in 2003.} This trend implies that Internet users are slowly increasing their fondness towards Unix-based OSs. There are three potential reasons for the aforementioned user affinity: 
\begin{itemize}
\item Various types of popular applications that were previously not supported by Unix-based OSs are nowadays being supported by the same with high ease of usability. For example, the highly popular text editor, \emph{Microsoft Word}, traditionally designed to run on Windows now has a MacOS version which is very easy to use. 
\item In addition to Unix-based OSs being more latently secure than Windows and increasingly supporting user popular applications, their GUIs are getting increasingly attractive and user friendly. 
\item Changing between ease of use OSs is a natural behavioral process which poses few technical, behavioral, or economic constraints on the user. 
\end{itemize}
Thus, we qualitatively conclude that OS platform switching (ex., from Windows to Unix-based OSs), or adopting secure OSs would enhance overall network security and favor the success of cyber-insurance markets, as it would mitigate the information asymmetry problem and reduce the interdependence and correlation of distributed system risks, without violating behavioral, economic, and policy constraints beyond a considerable extent. \emph{A thing to note here is that malicious users would want to spend more efforts in breaking into Unix-based OSs as soon as they start becoming popular, however there would then be a new OS which would be significantly better than Unix-based OSs in terms of latent security strength, thanks to parallel research developments in the field of operating systems. Thus, we firmly believe that irrespective of time, there would always be the scope of adopting a more secure OS. }
\subsubsection{Research Motivation}
With respect to OS switching dynamics it is evident that a cyber-insurer would prefer to insure a network of users having high willingness to adopt secure OSs in the long run. It is also important for the cyber-insurer to know how fast  would its clients settle into an invariant willingness value. The existence of willingness invariance and the speed\footnote{Why is the speed factor important? - because a high speed invariant settlement implies the practicality of designing appropriate cyber-insurance schemes. On the other hand if the invariant settlement time is large, it may contribute to the infeasibility of insurance schemes.} at which it could be achieved in turn depends on the network structure. We investigate on this important aspect in our paper. It is also intuitive that a cyber-insurer would want to incentivize (through its contracts) certain of its clients who are instrumental in the process of OS switching. In this regard we investigate theoretical properties of networks preferable to cyber-insurers, as well as mathematically identify users that have a significant impact on the process of OS switching.
\subsection{Our Contributions}
In this paper, we answer the following important questions - (i) how can we effectively model the OS platform switching process? (See Section II.), (ii) can we comment on how fast a distributed network of users would reach willingness invariance, if it exists?, for a given network, what are the upper bounds in the long run of individual user willingness to adopt secure OSs? what are the theoretical properties of networks a cyber-insurer prefers to insure based on willingness convergence speed and willingness upper bounds? (See Section III.), and (iv) how could a cyber-insurer incentivize/reward a client based on his topological location? (See Section IV.)

Through our contributions we mathematically prove that a cyber-insurer should prefer to insure networks where users on average have a high willingness to adopt secure OSs, and that the insurer should reward each user (via insurance contracts) in proportion to the impact they have on the average willingness degree of the whole network. Doing so would correlate to better self-defense investments by users, alleviating information asymmetry, in turn leading to successful cyber-insurance markets.


\section{Gossip-Based Switching Model}
In this section we propose a gossip-based switching model that captures the dynamics of OS platform switching in a distributed communication network. We first describe the model environment and follow it up with the description of our assumptions. We then mathematically capture the evolution of OS switching amongst users in the network. Finally, we study the convergence of the switching process. \\
\emph{Why a gossip-model?:} We adopt a gossip-based model because when it comes to embracing OSs, general Internet users tend to rely on public opinion to a considerable extent before deciding upon an OS to adopt for long term use. In this regard, there are social contagion and social influence models in theory as well, but we leave it to a later discussion (See Section V) of why we choose a gossip-based model over others. 
\subsection{Model Environment}
We consider a distributed communication network\footnote{The network could be the Internet or any other network having a decentralized communication system. However, from a cyber-insurance perspective, the network would most likely be the Internet or a part of it.} consisting of $n$ users. Each user adopts\footnote{We assume here that a user has only one computing device.} either a popular Windows operating systems (ex., XP, Vista), or one of Unix-based OSs such as Linux, MacOS, etc. as his primary\footnote{A user might have partitions on his computing device to store multiple OSs.} computing platform. Irrespective of his adopted OS, each user $i$ has a particular \emph{initial willingness} $w_{i}(0)\,\epsilon\,\mathbb{R}$ towards adopting a Unix-based operating system for the purposes of computing\footnote{We emphasize here that our decision to consider Unix-based OSs as latently more secure OSs than Windows OSs arise from the study in \cite{ramr}.}. This assumption is valid as many Windows users have the inclination to use Unix-based OSs but do not use it due to factors like confidence of operation, and incomplete information on Unix/Linux application support. We assume that $w_{i}()$ lies in the interval $[0,1]$. A user already adopting an Unix-based OS has a willingness value of 1. Each user in the distributed network updates his willingness values over time by interacting with his peers via an \emph{acquaintance network} (AN), which is an overlay network on top of the physical distributed network. The interactions could be through mutual physical contact or through online communication, or both. We assume here that the acquaintance network and the initial willingness values are known to the entity that takes decisions on cyber-insurance contracts. It could be a cyber-insurance agency, the government, or other policy makers. 

We use the asynchronous continuous-time model in \cite{bgsp} to represent interactions between users in the AN. In particular, each user interacts with other users at instances defined by a Poisson process of rate one, \emph{independent} of other users. Equivalently, interaction instances across all users occur according to a rate $n$ Poisson process at times $t_{r}$, $r \ge 1$. In order to entail simplicity of visualizing user interactions, there are $n$ time slots per absolute time \cite{bgsp}, where each time slot $r$ is of the discrete form $[t_{r}, t_{r +1})$. There is at most one meeting (interaction) instance per time slot, and therefore on average there are $n$ interaction instances per absolute time. 

Regarding an interaction, user $i$ meets (interacts with) user $j$ with probability $p_{ij}$. Upon interaction, users $i$ and $j$ update their individual willingness according to the following three possibilities. 
\begin{enumerate}
\item Each user updates his willingness as the average of each other's pre-meeting willingness as per the following relation.
\[w_{i}(r + 1) = w_{j}(r + 1) = \frac{w_{i}(r) + w_{j}(r)}{2}.\]
In this case users $i$ and $j$ are \emph{regular} with respect to each other. The probability of this type of meeting happening between $i$ and $j$ is $y_{ij}$. 
\item User $j$ influences user $i$, in which case for some $\delta\,\epsilon\,(0, \frac{1}{2}]\footnote{For a value of $\frac{1}{2}$, $i$ perceives $j$ as a regular user.}$, individual user willingness changes as per the following relations. 
\[w_{i}(r + 1) = \delta w_{i}(r) + (1 - \delta)w_{j}(r).\]
\[w_{j}(r + 1) = w_{j}(r).\]
In this case user $j$ is considered as an \emph{influential} user who influences $i$ towards adopting his preferred OS, but himself does not change his viewpoint\footnote{An analogous example would be of a mobile operator promoting a type of phone to its clients, i.e., it provides service on a particular phone type ( ex., say phones with Android OSs) and tries to influence potential clients to buy phones of the corresponding type.}. The probability of this type of meeting happening between $i$ and $j$ is $x_{ij}$. Examples of influential users include security experts, commercial organizations promoting a certain OS, etc.,   
\item Neither $i$ nor $j$ change their opinions on interaction, which imply the following.
\[w_{i}(r + 1) = w_{i}(r).\]
\[w_{j}(r + 1) = w_{j}(r).\]
In this case users $i$ and $j$ are said to be \emph{persistent} with respect to each other. The probability of this type of meeting happening between $i$ and $j$ is $z_{ij}$, where $z_{ij} = 1 - x_{ij} - y_{ij}$. 
\end{enumerate}
\emph{Important Note:} The entire analysis in this paper is related to the acquaintance network. Thus, when we speak of the right network a cyber-insurer is willing to target, we are referring to the right AN. The AN is also a distributed communication network but an overlay one. The results of our analysis on ANs can be mapped to the physical distributed network, i.e., the insurer prefers to insure its clients if the client AN graph satisfies certain properties, and it also incentivizes certain clients based on their topological position in the AN.
\subsection{Assumptions}r
We make the following assumptions in this paper.  
\begin{enumerate}
\item $p_{ii} = 0\,\forall i$, $p_{ij} \ge 0\,\forall i,j$, and $\sum_{j = 1}^{n}p_{ij} = \,\forall i$. 
\item The acquaintance network $AN = (V,E)$ is strongly connected, i.e., for all $i, j\,\epsilon\,V$, there exists a path connecting $i$ and $j$. A link $(a,b)$ in set $E$ gets formed if and only if $p_{ab} > 0$. This assumption implies that each user is socially accessible to any other user in the acquaintance network possibly through multiple links. In this regard let $d = max_{i,j\,\epsilon\,V}d_{ij}$ denote the maximum shortest path length between any $i,j\,\epsilon\,V$. 
\item For all $(i,j)\,\epsilon\,E$ in \emph{AN}, $x_{ij} + y_{ij} > 0$. This assumption states that irrespective of the type of user, each user gets influenced by some of his acquaintances. This assumption is realistic as even experts and organizations are prone to changing their views based on social influence. 
\end{enumerate}
\subsection{The Evolution of User Willingness}
We consider a vector $w(r) = (w_{1}(r), ........, w_{n}(r))$ denoting the vector of user willingness for Unix-based OSs at time slot $r$. The user willingness in time slot $r + 1$ is updated according to the following equation.
\begin{equation}
w(r + 1) = W(r)w(r),
\end{equation}
where $W(r)$ is stochastic random matrix\footnote{$W(r)$ is a matrix induced by the acquaintance network and the user interactions on it, for time slot $r$. The expression for $W(r)$ can be derived by applying simple concepts from \emph{random matrix theory}.} for all $r$ given by the following 
\begin{equation}
W(r) = \left\{
\begin{array}{rl}
X_{ij} \equiv I - \frac{(e_{i} - e_{j})(e_{i} - e_{j})^{T}}{2} & \text{w.p.}  \frac{p_{ij}y_{ij}}{n},\\
Y1_{ij} \equiv I - (1 - \delta)e_{i}(e_{i} - e_{j})^{T} & \text{w.p.}  \frac{p_{ij}x_{ij}}{n},\\
Z_{ij} \equiv I &\text{w.p.} \frac{p_{ij}z_{ij}}{n},
\end{array}\right.
\end{equation}
for all $i,j\,\epsilon\,V$, where $e_{i}$ is an n-dimensional vector with a 1 in position $i$ and zeros in other locations; $e^{T}$ is the transpose of vector $e$. The equation implies that $W(r)$ is a stochastic matrix\footnote{A stochastic matrix has each of its rows sum to 1.} for all $r$ and is also independent and identically distributed over all $r$. We now introduce transition matrices $\Psi$ of the following form. 
\begin{equation}
\Psi(r, s) = W(r)W(r - 1)...................W(s + 1)W(s), \forall r, s, r \ge s.
\end{equation}
We can rewrite Equation 1 using transition matrices in the following manner.
\begin{equation}
w_{i}(r + 1) = \sum_{j = 1}^{n}[\Psi(r,s)]_{ij}w_{j}(s).
\end{equation}
Since $W(r)$ is a random variable, we have 
\begin{equation}
E[W(r)] = \bar{W}, \, \forall r \ge 0,
\end{equation}
where $\bar{W}$ is the mean acquaintance matrix. We can write $\bar{W}$ as 
\begin{equation}
\bar{W} = \frac{1}{n}\sum_{i,j}p_{ij}[y_{ij}X_{ij} + x_{ij}Y_{ij} + z_{ij}I.]
\end{equation}
or 
\[\bar{W} = \frac{1}{n}\sum_{i,j}p_{ij}[(1 - z_{ij})X_{ij} + z_{ij}I] + \frac{1}{n}\sum_{ij}p_{ij}x_{ij}[Y_{ij} - X_{ij}].\]
We can express $\bar{W}$ as $\bar{W} = K + L$, where 
\[K = \frac{1}{n}\sum_{i,j}p_{ij}[(1 - z_{ij})X_{ij} + z_{ij}I].\]
and 
\[L = \frac{1}{n}\sum_{ij}p_{ij}x_{ij}[Y_{ij} - X_{ij}].\]
We note that $K$ only depends on the user meeting probabilities $p_{ij}$ of users and the probabilities $z_{ij}$ denoting the likelihood of two users, persistent relative to each other, meeting. Thus, $K$ can be considered as a \emph{social matrix} representing underlying social interactions amongst users, where the matrix is symmetric and doubly stochastic. i.e., both the row sums and column sums of the matrix sum to 1. This property would be useful in considering $K$ as a Markov chain while analyzing our model. The matrix $L$ on the other hand can be thought of as representing the influence structure because it  includes the influencing probabilities $x_{ij}$s and $y_{ij}$s. $L$ can be thought of as an \emph{influence matrix} that incorporates information about which users and links are influential and regular. \emph{We emphasize here that both $K$ and $L$  are based on the dynamics of social interactions on the AN.} 
\subsection{Convergence of User Willingness}
In this section we study the convergence of the OS platform switching process. We show that despite the presence of influential users existing in an acquaintance network, each with different initial willingness values, every user willingness value converges to a single common value in the long run. This implies that ultimately every user has the same willingness to switch to a secure OS/s. Note that this \emph{does not mean} that in the long-run every user will switch to a secure OS. We state the following theorems to highlight our results on willingness convergence. The proofs of the theorems are presented in the Appendix. \\ \\
\textbf{Theorem 1.} \emph{The user willingness sequence $\{w_{i}(r)\}$ converges to a common willingness value $\bar{w}$, such that}
\[lim_{r\rightarrow\infty}w_{i}(r) = \bar{w},\, \forall i\,\epsilon\,V, w.p. 1\]  
and 
\[\bar{w} = \sum_{j = 1}^{n}\pi_{j}w_{j}(0), \,\sum_{j = 1}^{n}\pi_{j} = 1, \pi_{j} \ge 0\,\forall j\]
\emph{where $\bar{w}$ is a limiting random variable.} \\ \\
\emph{Theorem Implications: }The Internet users in the long run reach a common agreement regarding the willingness to adopt a secure OS, i.e., each user has the same value of willingness to adopt secure OSs. The theorem result is interesting and counter-intuitive\footnote{It is strange to believe that each user will have the same willingness in the long run given that each might start the evolution process with widely varying willingness values. It seems more likely that there would be a cluster of users having different willingness values.} but makes sense in the light of the assumption that even persistent agents are prone to changing their willingness under influence of social interactions. Thus, in the long run each user willingness converges to a single random variable taking up a common value. It is evident that higher the common value, more preferable it becomes for a cyber-insurer to insure its clients. The common value depends on the order in which user acquaintances have taken place. This leads us to investigating the expected value of the random variable, and that is what we accomplish via the next theorem.\\ \\
\textbf{Theorem 2.} \emph{Let $\bar{w}$ be the limiting random variable of the willingness sequences $\{w_{i}(r)\},\, \forall i\,\epsilon\,V$. We have} 
\[lim_{r\rightarrow\infty}\bar{W}^{k} = e\bar{\pi}^{T}, \,\bar{\pi}^{T} = E[\pi].\]
and
\[E[\bar{w}] = \sum_{i = 1}^{n}\bar{\pi}w_{i}(0) = \bar{\pi}^{T}w(0),\]
\emph{where $e$ is the vector of all ones, and $\bar{\pi}$ is the stationary distribution w.r.t. mean acquaintance matrix $\bar{W}$.}\\ \\ 
\emph{Theorem Implications:} We note here that $\bar{\pi}$ is the stationary distribution indicating the weight given to each user in affecting the expected value of common converged value of willingness. This weight can be considered as a measure of the influence of each user on the average willingness in the long run. We will discuss more on user influence in subsequent sections, and its implications to cyber-insurance. An interesting thing to note is that in the absence of influential users, $x_{ij}$ is zero for all $i$ and $j$, and as a result the average value of $w_{i}(r)$ remains constant in each time slot $r$, and is equal to the average of the initial user willingness, and each user has the same amount of influence. The stationary distribution in this case is a $n$-dimensional vector where each element is $\frac{1}{n}$. 

\section{Willingness Bounds and its Importance to Cyber-insurers}
In the previous section, we showed that users in a network converge to a common willingness value in the long run. In this section we derive theoretical bounds on the performance of OS switching in Internet-like distributed systems. We define performance as an average function of the converged value of individual user willingness. In this regard we study how \emph{fast} a network of users converge to a common willingness value, and what are the properties of networks based on the speed of convergence and the value of the performance function. \emph{Our study in this section paves the way for cyber-insurers to make decisions on designing appropriate contracts for clients in different types of networks.}

\subsection{Performance Function}
We adopt the following performance function $P$ in this paper.
\begin{equation}
P(\bar{w}) = E[\bar{w} - \gamma] = E[\bar{w}] - \gamma = \sum_{i\epsilon V}(\bar{\pi} - \frac{1}{n})w_{i}(0),
\end{equation}
where 
\begin{equation}
\gamma = \frac{1}{n}\sum_{i = 1}^{n}w_{i}(0).
\end{equation}
\emph{Interpretation of Performance Function:} The performance function measures the deviation of the common converged willingness value from the average of the initial user willingness values. 
The rationale behind using such a function is the fact that we want to compare the limits of user influence by influential users with that when no influential users in the network are present. An important significance of the performance function is the impact of the presence of influential users in the network, which in turn correlates to the impact of individual user contributions to the common converged value. We will study more about individual influence contributions in Section IV, and its implications on cyber-insurance contracts. 

We state the following two theorems in relation to evaluating upper bounds of user willingness. The proofs of the theorems are in the Appendix.\\ \\
\textbf{Theorem 3.} (a) \emph{Let $\bar{\pi}$ denote the stationary distribution related to the common converged value of user willingness. Then the following relation holds.
\begin{equation}
||\bar{\pi} - \frac{1}{n}e||_{\infty} \le \frac{1}{1 - \rho}\frac{\sum_{i,j}p_{ij}x_{ij}}{2n},
\end{equation}
where $\rho$ is a constant equalling $(1 - n\Psi 1^{d})^{\frac{1}{d}}$, and $\Psi 1$ is given by the following 
\[\Psi 1= min_{(i,j)\epsilon E}\left\{\frac{1}{n}[p_{ij}\frac{1 - z_{ij}}{2} + p_{ji}\frac{1 - z_{ji}}{2}]\right\}.\]
(b) 
\begin{equation}
|E[\bar{w} -  \frac{1}{n}\sum_{i = 1}^{n}w_{i}(0)| \le \frac{1}{1 - \rho}\frac{\sum_{i,j}p_{ij}x_{ij}}{2n}||w(0)||_{\infty}
\end{equation}}\\ \\
\textbf{Theorem 4.} \emph{Let $\bar{\pi}$ denote the stationary distribution related to the common converged value of user willingness. Then the following relation holds.
\begin{equation}
||\bar{\pi} - \frac{1}{n}e||_{2} \le \frac{1}{1 - \lambda_{2}(K)}\frac{\sum_{i,j}p_{ij}x_{ij}}{n},
\end{equation}
where $\lambda_{2}(K)$ is the largest eigenvalue of the matrix $K$.}\\ \\
\emph{Implications of Theorems 3 and 4:} We jointly provide the implications of the two theorems as they are quite interrelated with each other. First, the two theorems provide \emph{upper bounds} on our performance function. Second, the theorem characterizes the variation of the stationary distribution in terms of (i) $\frac{\sum_{i,j}p_{ij}x_{ij}}{n}$, which is the average influence by influential users, and (ii) $\lambda_{2}(K)$, the second largest eigenvalue of the social matrix $K$. It is common knowledge that the term $1 - \lambda_{2}(K)$ is the spectral gap of a matrix \cite{pbr}  and controls the rate of convergence of the Markov chain induced by matrix $K$\footnote{Since $K$ is a stochastic matrix, it can be considered as a transition matrix of Markov chain.} to its stationary distribution $\bar{\pi}$. In theory, the larger the spectral gap, faster does powers of a Markov chain converge to its stationary distribution. A fast converging Markov chain is called \emph{fast-mixing}. Now Theorem 4 states that for a fast-mixing Markov chain induced by the social matrix, it would fast converge to the average of the initial willingness of individual users, i.e., $\gamma$, which in turn implies that the impact of influential users on other network\footnote{Here the term `network' implies the acquaintance network. In specific it refers to the network induced by the social matrix $K$, which is formed by user acquaintance dynamics on the AN.} users would be negligible with respect to making them adopt secure OSs. This is intuitive because in  a fast-mixing graph (Markov chain) there are several connections between any pair of users. In this regard, consider an influential user who  is influenced by some other users in the network. These users are in turn connected to other users in the network due to the fast-mixing property of the graph. Assuming that the number of non-influential users are much greater in number than influential users, the common converged willingness value for all users in a fast-mixing graph is the average of initial individual user willingness. \emph{Thus, a cyber-insurer would NOT prefer to insure its clients that form a fast-mixing social graph, OR they would like to insure the users via a different type of cyber-insurance contract than what they plan for users in a slow-mixing graph.} 

In a slow mixing graph (Markov chain) there is a high degree of clustering of influential users and these users get their influence from mostly regular users whom they have already influenced in the past. Due to the loosely connected clusters, the influential agents will spread their influence widely before they get feedback from users they already influenced in the past. As a result the commonly converged willingness value will be much closer to the mean willingness of influential users. \emph{Thus, a cyber-insurer would prefer to insure its clients that form a network inducing a slow-mixing Markov chain on its social graph.} 

\section{On Proportionally Incentivizing Clients}
In this section we focus on individual client effects on the common converged willingness value. The individual impacts provide cyber-insurers with ways to incentivize clients differently based on their amount of impact on the network performance function. In this paper we measure the impact of each individual user via the expression $\bar{\pi} - \frac{1}{n}$. The term $\frac{1}{n}$ comes from the average of all initial individual willingness and through individual impact we measure the deviation from the average of long-run individual willingness. The incentives would most likely be in the form of \emph{differential cyber-insurance contracts} with lesser premiums charged to network users who have a greater impact on the value of the network performance function. In this section we divide our analysis of individual impacts into three parts based on the effect of influential edges (between the influencer and the influencee) on graph connectivity: in the first part we consider general graphs induced by the social matrix where there may or may not be edges that are formed between an influencer and the influencee - we term these graphs as \emph{Class I graphs}, in the second part we consider graphs induced by a social matrix where edges formed between influential users and their influencees are necessary for the graph to be connected - we term these graphs as \emph{Class II graphs}, and finally we consider graphs in which edges formed between influential users and influencees are not necessary for the graph to be connected - we term these graphs as \emph{Class III graphs}. By the term `graphs' we imply graphs induced by the social matrix $K$, which in turn is formed via user interactions on the AN. Our analysis is heavily based upon the concept of mean first passage times in the theory of Markov chains. \emph{The rationale behind adopting this concept is that the impact of opinions is dependent on the social distance between users, and the passage times reflect the speed with which opinions reach users separated by certain social distances.} 

\subsection{Class I Graphs}
General graphs are induced by $K$, with the condition that there are no constraints on the presence of links between the influencer and the influencee. We provide our result on the amount of individual user impact in the form of the following theorem. The proof of the theorem is mentioned in the Appendix. \\ \\
\textbf{Theorem 5.} \emph{Let $\bar{\pi}$ denote the stationary distribution related to the common converged value of user willingness. Then the following relation holds for every user $k$. 
\begin{equation}
\bar{\pi_{k}} - \frac{1}{n} = \frac{1}{2n^{2}}\sum_{i,j}p_{ij}x_{ij}\left((1 - 2\delta)\bar{\pi_{i}} + \bar{p_{ij}}\right)(m_{ik} - m_{jk}),
\end{equation}
where $m_{ik}$ and $m_{jk}$ are the mean first passage times\footnote{The mean first passage time from node $i$ to node $j$ in a Markov chain having transition probability matrix $K$ is given as $m_{ij}$ and equals $\frac{Y_{jj} - Y_{ij}}{\pi_{j}}$, where $Y = \sum_{k=0}^{\infty}(K^{k} - K^{\infty})$ \cite{pbr} is the fundamental matrix of the Markov chain induced by $K$.}  from $i$ to $k$ and $j$ to $k$ respectively, in the Markov chain induced by $K$.}\\ \\
\emph{Theorem Implications: } The theorem provides an exact closed form expression for the impact of a user $k$ on the value of the network performance function in terms of the mean first passage times  from $k$ to the influential and influenced users. To provide an intuition, consider a single edge $(j,i)$ between influencer $j$ and influencee $i$. Thus, for $k\,\ne\,\{j,i\}$, its influence could be indirect on the willingness of $j$. In this regard $m_{jk}$ represents the distance between $j$ and $k$ and enters negatively into Equation 12. On the other hand any user who meets with user $i$ with a high probability would be influenced indirectly by $j$. Thus, the impact of user $k$ on the performance function would be increasing in $m_{ik}$, i.e., when $m_{ik}$ is small $k$ has negative impact as he is very closed to influenced agent $i$, whereas when $m_{jk}$ is small $k$ will have a positive impact on the performance function as his opinions would be quickly absorbed by $j$. The theorem generalizes our intention to multiple links of the form $(j,i)$. In the case of the absence any links of the form (influencer, influencee), the impact of $k$ is zero. 
\subsection{Class II Graphs}
We deal with graphs induced by $K$ where there exists links between influencers and the influencees, which when removed from the graph disconnects the graph. In this regard, we provide our result on the amount of individual user impact in the form of the following theorem. The proof of the theorem is mentioned in the Appendix. \\ \\
\textbf{Theorem 6.} \emph{Let $\bar{\pi}$ denote the stationary distribution related to the common converged value of user willingness.  Let there be an edge $(i,j)$ in the graph induced by $K$ such that $x_{ij} > 0$. Assuming that removing $(i,j)$ from  the graph would disconnect it, the following relation holds for all user $k$. 
\begin{equation}
\bar{\pi_{k}} - \frac{1}{n} = \frac{2}{n^{2}}\frac{\mu_{ij}(1 - \delta)}{1 - \frac{\mu_{ij}}{n}((1 + 2\delta)|N(i,j)| - |N(j,i)|)}\Omega_{ij}(k),
\end{equation}
where 
\[\mu_{ij} = \frac{p_{ij}x_{ij}}{p_{ij}(1 - z_{ij}) + p_{ji}(1 - z_{ji})},\]
and
\[\Omega_{ij}(k) = |N(i,j)|, k\,\epsilon\,N(j,i); \Omega_{ij}(k) = -|N(j,i)|, k\,\epsilon\,N(i,j),\]
where $N(i,j)$ and $N(j,i)$ are two disjoint sets of nodes on the removal of edge $(i,j)$ from the graph.}\\ \\
\emph{Theorem Implications:} The theorem implies that if there is a single influential edge between two user clusters, then the impact of each user within the same clusters, on the network performance function, are equal. Intuitively, all users in a cluster in which there is the influential user will in the long run shape the willingness of the influential user through their individual impacts, and surprisingly it is the same amount of impact for all the users irrespective of whether the user is directly connected to the influential user. This is because in the limiting distribution, the common converged value will impact the willingness value of the influential user, and since there is only one such influential user, the impact of all other non-influential users are the same. 
 
\subsection{Class III Graphs}
We deal with graphs induced by $K$ where there exist links between influencers and the influencees, which when removed from the graph need not necessarily disconnect the graph. In this regard, we provide our result on the amount of individual user impact on the network performance function in the form of the following theorem. The proof of the theorem is mentioned in the Appendix. \\ \\
\textbf{Theorem 7.} \emph{Let $\bar{\pi}$ denote the stationary distribution related to the common converged value of user willingness. Then the following relation holds for all user $k$.
\begin{equation}
|\bar{\pi_{k}} - \frac{1}{n}| \le \sum_{i,j}\frac{2p_{ij}x_{ij}}{n}(\frac{1 + \log n}{\psi}),
\end{equation}
where $\psi$ is the conductance of the Markov chain with transition matrix given by $K$.}\\ \\
\emph{Implications of Theorem 7:} The conductance of a Markov chain \cite{pbr} is defined as follows.
\begin{equation}
\psi = inf_{A\subset V}\frac{Q(A, A^{c})}{\pi(A)\pi(A^{c})},
\end{equation}
where $Q(A,A^{c}) = \sum_{i\epsilon A, j\epsilon A^{c}}\pi_{i}K_{ij}$, and $\pi(A) = \sum_{i\epsilon A}\pi_{i}$. The conductance resembles the minimum probability that a Markov chain goes from a state in $A$ to a state in $A^{c}$. Thus, the conductance is an appropriate measure of the mixing time of the graph induced by social matrix $K$. Greater the conductance value, the more connected is the graph. The theorem gives the expression for the upper bound of the individual impact on the performance function for Class III graphs. It is evident that greater the conductance the impact decreases due to the graph getting better connected. 
\section{Related Work}
In this section, we briefly describe related work as applicable to our paper. We compare our work with three different research areas suited to influencing Internet users to adopt Unix-based OSs, viz., \emph{influence maximization}, \emph{diffusion of ideas}, and \emph{cascading effects due to cyber-insurance adoption}.

\subsection{Influence Maximization}
The problem of OS platform switching bears some resemblance to the problem of influence maximization in social networks, even though it is different than influence maximization problem. The latter problem was first studied by Domingos and Richardson in \cite{drr}, where a social network of potential customers of a market product is modeled as a Markov random field, and probabilistic techniques are used to find those customers to target to for effective \emph{viral marketing}\footnote{Viral marketing is a marketing technique used by companies to promote the cascading of new products or innovative ideas. The technique exploits the network value of customers in order to cascade the adoption of new ideas/products.}. In a seminal piece of work, Kempe et.al. \cite{kkt} study influence maximization in social networks via the following algorithmic problem - given a social network graph and influence probabilities on each edge, how do we select a small set of initial users so as to maximize the set of users who get influenced. The authors model the problem as a discrete optimization problem assuming suitable models of information diffusion\footnote{The paper considers the \emph{Linear Threshold Model} and the \emph{Independent Cascade Model} of diffusion.} They prove the problem to be NP-Hard and propose constant factor greedy approximable algorithms based on submodular functions \cite{fjir} that find the initial set that guarantee a solution that is within 63\% of the optimal. The authors also show that their algorithms out-perform node-selection heuristics based on the concepts of \emph{degree centrality} and \emph{distance centrality} used in social network analysis. Following up on the work by Kempe et.al., several works \cite{cwy2}\cite{cwy}\cite{cyz}\cite{ksrt}\cite{lah} have addressed the algorithmic version of the influence maximization problem and bettered the greedy algorithm proposed in \cite{kkt}. 

\emph{Drawbacks of existing solutions} - The related works on influence maximization are mainly targeted towards ways to \emph{effectively} market new products/novel ideas developed by an organization. On the other hand, our paper formulates the base or \emph{pre-step}, using which ways to effectively market cyber-insurance adoption can be designed. However, the existing works have the following implementability drawbacks on parameters common to our work:  (1) they only assume an influence value for each neighbor of a user and use the threshold model to judge whether a user would adopt an OS, but the works do not \emph{theoretically} model the time (dynamics of the influence process) it takes for users to reach willingness convergence. We model the time dynamics in this paper as a stochastic process. The analysis helps the cyber-insurer have an idea regarding the time feasibility of adoption, and (2) The influence of each potential product user on his neighbors in the social graph is assumed to be known to the organization. This is certainly not the case in reality and also in the case of an ISP (a potential cyber-insurance agency) willing to indulge its clients in using Unix-based OSs. An ISP may at best know the social graph but not the individual influence degree. Thus, the system model proposed in \cite{kkt} and subsequent related works in order to achieve influence maximization, is not implementation realistic\footnote{An ISP could design efficient incentive-compatible game-theoretic mechanisms to enable users to willingly and truthfully confide to the ISP their influence values on social contacts, but users might themselves not have a clear idea of the influence values in the first place.} 
\subsection{Diffusion of Ideas} 
Models from the theory of diffusion have been used in existing literature to model and explain the dynamics of adoption of new ideas in a social network. In some basic models \cite{ekr}\cite{vrr}, a user's decision to adopt a new idea is based on the decisions of its neighbors in the social graph. These models follow the principle that a user adopts a new idea only if a certain number of its neighbors in the social graph adopt the new idea, where individual user thresholds are assumed to be homogenous across the users. In a different type of a diffusion model introduced in \cite{ellison}\cite{morrisr}, the users adopt a new idea where the threshold value is a function of the payoff of a \emph{coordination game}. In a recent work \cite{lelarge}, the author uses the coordination game model to analyze the spread of new behaviors in a random social graph and show the following: (1) When the social network is sufficiently sparse, the contagion is limited by the low connectivity of the network; when it is sufficiently dense, the contagion is limited by the stability of the high-degree nodes. This phenomenon explains why contagion is possible only in a given range of the global connectivity, (2) When contagion is possible, both in the low and high-connectivity cases, the number of pivotal\footnote{the largest component of players requiring a single neighbor to change strategy in order to follow the change.} players is low, resulting in rare occurences of cascades. However in the high-connectivity case, the system displays a robust-yet-fragile quality: while the cascades are very rare, their sizes are very large. This feature makes global contagions exceptionally hard to anticipate, and (3) When the initial number of adopters of a new idea is small, the idea spreads for low global-connectivity, whereas high global connectivity inhibits global contagion, but once it occurs, the connectivity facilitates spread. 

\emph{Drawbacks of existing solutions:} The models in \cite{ekr}\cite{vrr} assume homogenous adoption thresholds for individual users. However in reality, each user has different thresholds\footnote{The Linear threshold model and the Independent cascade model do account for heterogenous thresholds.} In regard to the models based on the coordination game, each one of them assume equal payoffs for all users for adopting a given choice. However, in reality each user is most likely to have different payoffs for adopting a given choice. In addition, the models assume a zero payoff for two users when they coordinate and find that each has a different adoption choice. This again is unrealistic as each user is more likely to have an increase/decrease in his payoff on coordination resulting in contradictory opinions, rather than him landing directly on a zero payoff. The work in \cite{lelarge} base its analyses and results on a random social graph. However, random graphs hardly represent any real-life phenomena. 
\subsection{Cascading Effects Due to Cyber-Insurance}
Lelarge and Bolot in \cite{leb8} have modeled the dynamics of the Internet users investing in self-defense, and shown that cyber-insurance incentivizes Internet users to optimally invest in self-defense investments and helps cause a self-defense cascade, i.e., incentivizing a certain number of Internet users to invest in self-defense causes all the Internet users to invest in self-defense, in turn increasing overall network security. In deriving their results, the authors in \cite{leb8} account for the `externality' factor and the `free-riding' problem related to user security investments, i.e., security investments by a user's neighbors generates a positive externality for the user, and might result in the user investing sub-optimally in self-defense. To prove the occurrence of a self-defense cascade, the authors use utility-theoretic comparisons to show more benefit to a user due to him optimally investing in self-defense mechanisms, when compared to him not investing optimally.  

\emph{Drawbacks of the existing solution:} The authors analyze random graphs and base their results on such graphs, which do not represent real-life phenomena. In our work, we address arbitrary graphs rather than graphs of any special kind. In addition, they represent the Internet topology as a random graph, which is proven not to be the case \cite{fff}. However, it is not realistic to assume that the cascading phenomenon as mentioned in \cite{leb8} is the way cascading might occur in the Internet. This is because the model in \cite{leb8} does not (i) model facets of human behavior (influence, coordination) that are important for practically achieving cascades and (ii) account for the fact that a node neighbor (In the Internet topology graph) may not have any social relationship with the node to share influence on the factor of investing in self-defense.  
\subsection{Difference Between Our Problem and Existing Works}
Most related works study new-idea cascades as some form of a `diffusion of adoption' problem. In our work, we solve a problem similar to the diffusion problem in \cite{leb8}.  However, we do not aim to study ``diffusion of adoption" as in \cite{leb8} and other related works, but investigate the ``diffusion of willingness" problem because of the following two reasons: 
\begin{itemize}
\item In reality, the decision of whether an Internet user $i$ will switch his OS depends on the evolution of $i$'s change willingness \emph{(a psychological factor)}  over time. Higher the willingness of a user to change his OS in the long-run, greater are his chances of actually doing so. The willingness is driven by individual factors such as ease of OS use, application support, etc., as well as influences from social contacts. The model in \cite{leb8} is not practical, specifically in the case of switching OSs because it is simply \emph{may not} be the case that a user changes OS if majority of his neighbors in the social graph use a different OS than he uses; the user's willingness to change OS also depends on the influence that these neighbors (and possibly some non-neighbors) exert and his own personal want. As an example, it may well be the case that all of a Windows user's neighbors use Unix-based OSs but the user is stubborn enough to not being influenced by any of his neighbors. Our assumptions provide an explanation of why the rate of users changing from Windows to Unix-based OSs today, is so slow - the fact is that some users just do not want to stop using the Windows OS. 
\item By solving the ``diffusion of willingness" problem via relaxing some impractical modeling and topological assumptions made in \cite{leb8} and \cite{lelarge}, we plan to answer the following important question: \emph{Is there hope to mitigate the information asymmetry problem in cyber-insurance through platform switching, and thus enable the successful existence of cyber-insurance markets?}
\end{itemize}
\section{Conclusion}
In this paper, we argue in favor of OS platform switching (towards secure OSs) to be a way to enforce successful cyber-insurance markets. In this regard we have studied the dynamics of the OS switching process amongst users in an acquaintance network, which is an overlay network over a physical distributed communication network. Our analysis heavily relied on the theory of Markov chains. We found that cyber-insurers would prefer to insure users in a slow-mixing social graph due to high performance on such graphs w.r.t. final averaged willingness of users to adopt secure OSs.  For fast-mixing graphs, the cyber-insurers would have to design contracts via effective mechanism design to entail successful markets. We also proved upper bounds on the performance function in a given social graph induced by an acquaintance network. We showed that the upper bound is higher in case of slow-mixing graphs when compared to high-mixing graphs. Finally, we computed exact expressions for the impact of each individual user in an acquaintance network on the final converged value of willingness of users. Based on the impact value the cyber-insurer would differentiate contracts, charging inexpensive premiums for high impact users and higher premiums for low impact users.

\bibliographystyle{plain}
\bibliography{alluvion12}
\section{Appendix}
In this section we provide proofs of the theorems stated in the paper. \\ \\
\emph{Proof of Theorem 1.} Let $M(r) = max_{i\epsilon V}w_{i}(r)$ and let $m(r) = min_{i\epsilon V}w_{i}(r)$. For any $r\ge 0$ we obtain the following
\begin{equation}
E[M(r) - m(r)] \le [C (M(0) - m(0))],
\end{equation}
where $C = [1 - (\frac{\eta^{d}}{2})^{n^{2}} +  (\frac{\eta^{d}}{2})^{n^{2}}(1 - \frac{n\eta^{d}}{2\delta^{n^{2} - 1}})]^{\lfloor\frac{r}{n^{2}d}\rfloor}$. This implies that 
\[lim_{r\rightarrow\infty}M(r) - m(r) = 0,\,\,w.p.1.\]
The stochasticity of matrix $W(r)$ implies that sequences $\{M(r)\}$ and $\{m(r)\}$ are bounded and monotone and therefore converges to the same limit $\bar{w}$. Thus, $lim_{r\rightarrow\infty} w_{i}(r) = \bar{w}$. Now let $s = 0$. Thus, for all $i$ we have 
\begin{equation}
w_{i}(r) = \sum_{i,j}^{n}[\Psi(r - 1, 0)]_{ij}w_{j}(0),\forall r \ge 0
\end{equation}
Now for any initial willingness vector $w(0)$, the limit $lim_{r\rightarrow\infty}w_{i}(r)$ exists and is independent of $i$. Thus, for any $h$, the limit $lim_{r\rightarrow\infty}[\Psi(r - 1, 0)]_{ih}$ exists and is independent of $i$. Denoting this limit as $\pi_{h}$, and using above equations we prove the second part of our theorem result. $\blacksquare$\\ \\
\emph{Proof of Theorem 2.} The first part of the theorem follows from Theorem 4.1.4 in \cite{kesn}. For the second part of the theorem we have for all $r\ge 0$
\[w(r) = \Psi(r - 1, 0)w(0).\]
Since $w(r)\rightarrow \bar{w}e$ in the limiting case of $r$ being $\infty$, we have the following result in the light of the Lebesgue Dominated Convergence Theorem \cite{rudin}. 
\begin{equation}
E[\bar{w}e] = E[lim_{k\rightarrow\infty }w(r)] = lim_{r\rightarrow\infty}E[w(r)].
\end{equation}
Under the assumption that matrices $W(r)$ are independent and identically distributed over all $r$, we have 
\begin{equation}
E[\bar{w}e] =  lim_{r\rightarrow\infty}E[\Psi(r-1, 0)w(0)] = lim_{r\rightarrow\infty}\bar{W}^{r}(0),
\end{equation}
which in turn implies $E[\bar{w}] = \bar{\pi}^{T}w(0)$, thus proving the theorem. $\blacksquare$\\ \\
\emph{Proof of Theorem 3.}  For the first part of the theorem we can start by using a result from peturbation theory in Markov chains to the difference between $bar{\pi}$ and $\frac{1}{n}e$. According to the result, the following holds.
\begin{equation}
(\bar{pi} - \frac{1}{n}e)^{T} = \frac{1}{n}e^{T}LY(I - LY)^{-1},
\end{equation}
where $Y$ as mentioned previously is the fundamental matrix of $K$. The equation further evaluates to 
\begin{equation}
||\bar{\pi} - \frac{1}{n}e||_{\infty} \le ||LY||_{\infty}
\end{equation}
We now find an upper bound for $||LY||_{\infty}$, where $LY = \sum_{r = 0}^{\infty}LK^{r}$. From the linear update rule we have for any $a\epsilon\mathbb{R}^{n}$, $LK^{r}a(0) = La(r)$ for all $r$. Thus, the following relation is achieved.
\begin{equation}
LK^{r}a(0) = \frac{1}{n}\sum_{i,j}p_{ij}x_{ij}a^{ij}(r),
\end{equation}
where $a^{ij}(r)$ is equal to $Y1_{ij} - X_{ij}]a(r),\,\forall i,j,k\ge0$. Now we have 
\begin{equation}
||LK^{r}a(0)||_{\infty} \le \frac{1}{2n}(\sum_{i,j}p_{ij}x_{ij})\rho^{r}(M(0) - m(0)).
\end{equation}
Since $M(0) - m(0) \le 1$, we have 
\begin{equation}
||LYa(0)||_{\infty} \le \frac{\sum_{i.j}p_{ij}x_{ij}}{2n(1 - \rho)}
\end{equation}
We thus get 
\[||\bar{\pi} - \frac{1}{n}e||_{\infty} \le \frac{1}{1 - \rho}\frac{\sum_{i,j}p_{ij}x_{ij}}{2n}\]
For the second part of the theorem, we have $E[\bar{w}] = \bar{pi}w(0)$. This implies the following.
\begin{equation}
|E[\bar{w}] - \frac{1}{n}\sum_{i = 1}^{n}w(0)| = |\bar{\pi}^{T}w(0) - \frac{1}{n}e^{T}w(0)|\le||\bar{\pi} - \frac{1}{n}e||_{\infty}||w(0)||_{\infty}.
\end{equation}
This equation in conjunction with the result in the first part of the theorem proves the theorem. $\blacksquare$\\ \\
\emph{Proof of Theorem 4.} We know that 
\begin{equation}
||\bar{\pi} - \frac{1}{n}e||_{2} \le ||LY||_{2}.
\end{equation}
We focus on finding the upper bound of $||LY||_{2}$.  Let $a(0)\,\epsilon\,\mathbb{R}^{n}$ be an initial vector with $||a(0)||_{2} = 1$ and let there be the following sequence
\[a(r + 1) = Kx(r)\,\forall r \ge 0.\]
Then for all $r$ we have the following equation. 
\begin{equation}
LK^{k}a(0) = \frac{1}{n}\sum_{i,j}p_{ij}x_{ij}a^{ij}(r)
\end{equation}  
The upper bound of $||a^{ij}(r)||^{2}_{2}$ is computed as $||a(r) - \bar{a}e||^{2}_{2}$. Now note that $a(r) - \bar{a}e$ is orthogonal to $e$, which is the eigen vector corresponding to the largest eigenvalue $\lambda_{1} = 1$ of matrix $K$. Thus we have the following relation.
\begin{equation}
||a(r) - \bar{a}e||_{2}^{2}|| \le (\lambda_{2}(K)^{2})^{r}||a(0) - \bar{a}e||^{2}_{2} \le \lambda_{2}(K)^{2r},
\end{equation}
where $\lambda_{2}(K)$ is the second largest eigenvalue of matrix $K$.  Thus, $||a^{ij}(r)||_{2} \le \lambda_{2}(K)^{r},\,\forall r\ge 0$. Using the result of the fundamental matrix $Y$, we get 
\begin{equation}
||LYa(0)||_{2} \le \frac{1}{1 - \lambda_{2}(K)}\frac{\sum_{i,j}p_{ij}x_{ij}}{n},
\end{equation}
for any vector $a(0)$ with $||a(0)||_{2} = 1$. Combining this result with Equation 16, we prove the theorem. $\blacksquare$ \\ \\
\emph{Proof of Theorem 5.} Based on the Markov chain theory we can write the individual impact of user $k$ as 
\begin{equation}
\bar{\pi_{k}} - \frac{1}{n} = \bar{\pi}^{T}L[Y]^{k},
\end{equation}
where $Y$ is the fundamental matrix of the Markov chain with transition matrix $K$ and $L$ equals $ \frac{1}{n}\sum_{ij}p_{ij}x_{ij}[Y_{ij} - X_{ij}]$.  Now $L[Y]$ can be expressed as 
\begin{equation}
[L[Y]^{k}]_{l} = \sum_{i,j}\frac{p_{ij}x_{ij}}{n}\left\{
\begin{array}{rl}
(\frac{1}{2} - \delta)(Y_{jk} - Y_{ik}) & if\, l = i,\\
\frac{1}{2}(Y_{jk} - Y_{ik}) & if\, l = j,\\
0 & otherwise.
\end{array}\right.
\end{equation}
Thus, we have the following equation from the above relationships. 
\begin{equation}
\bar{\pi_{k}} - \frac{1}{n} = \frac{1}{2n}\sum_{i,j}p_{ij}x_{ij}\left((1 - 2\delta)\bar{\pi_{i}} + \bar{p_{ij}}\right)(Y_{jk} - Y_{ik})
\end{equation}
Substituting the value of $Y_{jk} - Y_{ik}$ as $\frac{1}{n}(m_{ik} - m_{jk})$ into the above equation, we obtain our theorem result. $\blacksquare$ \\ \\
\emph{Proof of Theorem 6.} Since $K$ is a doubly stochastic matrix, we have $m_{ij} = \frac{|N_{i.j}|}{K_{ij}}$, for every $k\,\epsilon\,N(j,i), m_{ik} - m_{jk} = m_{ij}.$ Since $(i,j)$ is an influential 
link we have for every $k\,\epsilon\,N(j,i)$ the following relation 
\begin{equation}
m_{ik} - m_{jk} = \frac{|N(i,j)|}{K_{ij}} = \frac{2n|N(i,j)|}{p_{ij}(1 - z_{ij}) + p_{ji}(1 - z_{ji})}.
\end{equation}
Similarly for every $k\,\epsilon\,N(i,j)$ we get
\begin{equation}
m_{ik} - m_{jk}  = -\frac{|N(i,j)|}{K_{ij}}= -\frac{2n|N(i,j)|}{p_{ij}(1 - z_{ij}) + p_{ji}(1 - z_{ji})}.
\end{equation}
Using the preceding relations we can express the relative mean passage time as 
\begin{equation}
m_{ik} - m_{jk} = \frac{2n\mu_{ij}}{p_{ij}x_{ij}}\Omega_{ij}(k).
\end{equation}
Now since $(i,j)$ is the only influential link, we have 
\begin{equation}
\bar{\pi_{k}} - \frac{1}{n} = (\frac{2}{n^{2}})\frac{\mu_{ij}(1 - \delta)}{1 - \frac{\psi_{ij}}{n^{2}}}\Omega_{ij}(k),
\end{equation}
where 
\[\psi_{ij} = \frac{p_{ij}x_{ij}}{2}[(1 + 2\delta)m_{ij} - m_{ji}].\]
Combining the above results we prove Theorem 6. $\blacksquare$ \\ \\ 
\emph{Proof of Theorem 7.} For every $k$ we have
\[|\bar{\pi_{k}} - \frac{1}{n}| = \frac{1}{2n^{2}}\sum_{i,j}p_{ij}x_{ij}\left((1 - 2\delta)\bar{\pi_{i}} + \bar{p_{ij}}\right)|m_{ik} - m_{jk}|\]
\[\le \sum_{i,j}\frac{p_{ij}x_{ij}}{2n^{2}}|m_{ik} - m_{jk}| \le \sum_{i,j}\frac{p_{ij}x_{ij}}{2n^{2}}max\{m_{ik}, m_{ji}\}\]
Since $m_{ik} \le m_{ij} + m_{jk}$ and $m_{jk}\le m_{ji} + m_{ik}$, we have 
\begin{equation}
|\bar{\pi_{k}} - \frac{1}{n}| \le \sum_{i,j}\frac{p_{ij}x_{ij}}{2n^{2}}max\{m_{ik} + m_{ji}\}
\end{equation}
Applying the following relation from \cite{alfilr}, i.e.,
\[max_{i,j}\{m_{ij} + m_{ji}\} \le \frac{4(1 + \log n)}{\psi\, min\, \pi_{k}}\]
we get 
\[|\bar{\pi_{k}} - \frac{1}{n}| \le \sum_{i,j}\frac{2p_{ij}x_{ij}}{n}(\frac{1 + \log n}{\psi}). \quad \blacksquare\]



\end{document}